\documentstyle[12pt,epsf]{article}
\def\fo{\hbox{{1}\kern-.25em\hbox{l}}}
\def\fnote#1#2{\begingroup\def\thefootnote{#1}\footnote{#2}\addtocounter
{footnote}{-1}\endgroup}

\renewcommand{\thefootnote}{\fnsymbol{footnote}}

\def\beq{\begin{equation}}
\def\eeq{\end{equation}}
\def\eq{\end{equation}}
\def\to{\rightarrow}

\def\bsg{\ifmmode B\to X_s\gamma\else $B\to X_s\gamma$\fi}
\def\bsll{\ifmmode B\to X_s\ell^+\ell^-\else $B\to X_s\ell^+\ell^-$\fi}
\def\bstt{\ifmmode B\to X_s\tau^+\tau^-\else $B\to X_s\tau^+\tau^-$\fi}
\def\shat{\ifmmode \hat{s}\else $\hat{s}$\fi}

\def\EmissT{\not \! \!  E_{T}}

\newcommand{\newc}{\newcommand}

\newc{\lcal}{\int {\cal L}dt}
 
\newc{\LSP}{{\chi^0_1}}
\newc{\stauR}{{\tilde \tau_R}}
\newc{\stau}{{\tilde \tau_1}}
\newc{\mstop}{m_{\tilde{t}}}
\newc{\mHpm}{m_{H^\pm}}
\newc{\gsim}{\lower.7ex\hbox{$\;\stackrel{\textstyle>}{\sim}\;$}}
\newc{\lsim}{\lower.7ex\hbox{$\;\stackrel{\textstyle<}{\sim}\;$}}
\newc{\ie}{{\it i.e.}}          
\newc{\etal}{{\it et al.}}
\newc{\eg}{{\it e.g.}}          
\newc{\kev}{\hbox{\rm\,keV}}            
\newc{\mev}{\hbox{\rm\,MeV}}            
\newc{\gev}{\hbox{\rm\,GeV}}            
\newc{\tev}{\hbox{\rm\,TeV}}
\newc{\xpb}{\hbox{\rm\, pb}}
\newc{\xfb}{\hbox{\rm\, fb}}

\newc{\mtop}{m_t}
\newc{\mbot}{m_b}
\newc{\mz}{m_Z}
\newc{\mw}{M_W}
\newc{\alphasmz}{\alpha_s(m_Z^2)}
\newc{\swsq}{\sin^2\theta_W}
\newc{\tw}{\tan\theta_W}
\newc{\cw}{\cos\theta_W}
\newc{\sw}{\sin\theta_W}
\newc{\BR}{\hbox{\rm BR}}
\newc{\zbb}{Z\to b\bar}
\newc{\Gb}{\Gamma (Z\to b\bar b)}
\newc{\Gh}{\Gamma (Z\to \hbox{\rm hadrons})}
\newc{\rbsm}{R_b^\hbox{\rm sm}}
\newc{\rbsusy}{R_b^\hbox{\rm susy}}
\newc{\drb}{\delta R_b}

\newc{\sgn}{\mbox{sgn}}
\newc{\tbeta}{\tan\beta}
\newc{\uL}{{\tilde u_L}}
\newc{\uR}{{\tilde u_R}}
\newc{\cL}{{\tilde c_L}}
\newc{\cR}{{\tilde c_R}}
\newc{\tL}{{\tilde t_L}}
\newc{\tR}{{\tilde t_R}}
\newc{\dL}{{\tilde d_L}}
\newc{\dR}{{\tilde d_R}}
\newc{\sL}{{\tilde s_L}}
\newc{\sR}{{\tilde s_R}}
\newc{\bL}{{\tilde b_L}}
\newc{\bR}{{\tilde b_R}}
\newc{\eL}{{\tilde e_L}}
\newc{\eR}{{\tilde e_R}}
\newc{\mhp}{m_{H^\pm}}
\newc{\mhalf}{m_{1/2}}
\newc{\emt}{{e/\mu /\tau}}

\newc{\lR}{\tilde{l}_R}
\newc{\lL}{\tilde{l}_L}
\newc{\nL}{\tilde{\nu}_L}
\newc{\na}{\chi^0_1}
\newc{\nb}{\chi^0_2}
\newc{\nc}{\chi^0_3}
\newc{\nd}{\chi^0_4}
\newc{\ca}{\chi^{\pm}_1}
\newc{\cb}{\chi^{\pm}_2}
\newc{\camp}{\chi^\mp_1}
\newc{\cbmp}{\chi^\mp_1}
\newc{\capos}{\chi^{+}_1}
\newc{\caneg}{\chi^{-}_1}
\newc{\phit}{\phi_t}
\newc{\phib}{\phi_b}
\newc{\phiew}{\phi_{ew}}
\newc{\htz}{h^0_t}
\newc{\hbz}{h^0_b}
\newc{\hewz}{h^0_{ew}}
\newc{\hsmz}{h^0_{sm}}
\newc{\huz}{h^0_u}
\newc{\hsusyz}{h^0_{susy}}

\def\NPB#1#2#3{Nucl. Phys. B {\bf #1}, #3 (19#2)}
\def\PLB#1#2#3{Phys. Lett. B {\bf #1}, #3 (19#2)}

\def\PRD#1#2#3{Phys. Rev. D {\bf #1}, #3 (19#2)}
\def\PRL#1#2#3{Phys. Rev. Lett. {\bf#1}, #3 (19#2)}

\def\MPL#1#2#3{Mod. Phys. Lett. A {\bf #1} (19#2) #3}
\def\beq{\begin{equation}}
\def\eeq{\end{equation}}
\def\bea{\begin{eqnarray}}
\def\eea{\end{eqnarray}}
\def\slashchar#1{\setbox0=\hbox{$#1$}           
   \dimen0=\wd0                                 
   \setbox1=\hbox{/} \dimen1=\wd1               
   \ifdim\dimen0>\dimen1                        
      \rlap{\hbox to \dimen0{\hfil/\hfil}}      
      #1                                        
   \else                                        
      \rlap{\hbox to \dimen1{\hfil$#1$\hfil}}   
      /                                         
   \fi}                                         %
\catcode`@=11
\long\def\@caption#1[#2]#3{\par\addcontentsline{\csname
  ext@#1\endcsname}{#1}{\protect\numberline{\csname
  the#1\endcsname}{\ignorespaces #2}}\begingroup
    \small
    \@parboxrestore
    \@makecaption{\csname fnum@#1\endcsname}{\ignorespaces #3}\par
  \endgroup}
\catcode`@=12

\def\jfig#1#2#3{
 \begin{figure}
 \centering
 \epsfysize=3.0in
 \hspace*{0in}
 \epsffile{#2}
 \caption{#3}
 \label{#1}
 \end{figure}}

\begin{document}

\begin{titlepage}

\begin{flushleft}
\end{flushleft}
\begin{flushright}
SLAC-PUB-7788 \\
hep-ph/9804242\\
March 1998
\end{flushright}
\bigskip



\huge
\begin{center}
The importance of tau leptons for \\
supersymmetry searches at the Tevatron
\end{center}

\large

\vspace{.15in}
\begin{center}

James D.~Wells\fnote{\dagger}{Work 
supported by the Department of Energy
under contract DE-AC03-76SF00515 and DE-FG-05-87ER40319.} \\

\vspace{.1in}
{\it Stanford Linear Accelerator Center \\
Stanford University, Stanford, California 94309 \\}

\end{center}
 
 
\vspace{0.15in}
 
\begin{abstract}

Supersymmetry is perhaps most effectively probed at the Tevatron through
production and decay of weak gauginos.  Most of the analyses of weak
gaugino observables require electrons or muons in the final state.  However,
it is possible that the gauginos will decay primarily to $\tau$ leptons,
thus complicating the search for supersymmetry.  The motivating reasons for
high $\tau$ multiplicity final states are discussed in three approaches
to supersymmetry model building: minimal supergravity, minimal gauge
mediated supersymmetry, and more minimal supersymmetry.  The concept
of ``$e/\mu /\tau$ candidate'' is introduced, and an observable
with three $e/\mu /\tau$ 
candidates is defined in analog to the trilepton observable.
The maximum mass reach for supersymmetry is then estimated when
gaugino decays to $\tau$ leptons have full branching fraction.

\end{abstract}

\begin{center}

{\small
Presented at the D0 New Phenomena Workshop,
UC, Davis, 26-28 March 1998}

\end{center}

\end{titlepage}

\baselineskip=18pt



\vfill
\eject

\section{Introduction}
\bigskip

The effects of minimal supersymmetry on hadron collider observables
are not expected to be 
spectacular.  It would be easier if supersymmetry strongly
implied additional resonances in $p\bar p$ scattering, or narrow
invariant mass peaks of two leptons, etc.  Perhaps we will find high
multiplicity lepton, photon or jet signals from R-parity violation 
or prompt decays to gravitinos in low energy supersymmetry breaking
theories.  This is not the most likely scenario.
Simplicity, proton stability arguments, and dark matter considerations
constitute a mild preference for R-parity conservation.
Furthermore, it appears most
natural in gauge mediated models to have the superpartners feel only
a small fraction of the full supersymmetry breaking
in nature~\cite{dinenelson}.  
Thus, prompt decays of the NLSP (next lightest supersymmetric
partner) on the time scales of detector sizes are not 
favored~\cite{prompt}.

The classic signatures of supersymmetry are not quite as spectacular
as those created by promptly decaying NLSPs or R-parity violation.
Perhaps the most probing signal~\cite{trileptons,tev_trilepton} 
of supersymmetry at the Tevatron is
\beq
p\bar p\to \chi_1^\pm \chi_2^0\to 3l+\EmissT
\eeq
where $l=e$ or $\mu$.
Cutting on the $p_T$ values of these leptons and requiring that
$m_{l^+l^-}\neq m_Z$, leaves a small background from
$W^\pm Z$ production followed by $W\to l\nu$ and $Z\to \tau\tau$, where
the $\tau$'s decay leptonically to $e\nu\nu$ or $\mu\nu\nu$.
The maximum
mass reach of degenerate gaugino-like $\chi_1^\pm$ and $\chi_2^0$
states is about $200\gev$ with $2\xfb^{-1}$ of integrated 
luminosity~\cite{tevupgrade}.

One of the main purposes here is
to demonstrate that it is quite natural to
expect that the $3l$ signal is not present at any reasonable rate
at the Tevatron.  Instead, one expects over a large region of
supersymmetry parameter space to be dominated by multiple
$\tau$ events
from $\chi^+_1\chi^0_2$ production. 
The reasons for this are explained in the subsequent section
for three different approaches to supersymmetry
model building (standard supergravity scenarios, gauge mediated models,
and ``more minimal'' supersymmetry). Because $\tau$'s are more
difficult to tag and identify than leptons, and because quark and gluon jets
sometimes look like $\tau$'s, the searches in this mode are 
difficult.  Some estimates are made for the search
capabilities.

\section{Reasons for $3\tau$ events}
\medskip

There are several reasons why $\chi^\pm_1$ may decay preferentially
into $\tau^\pm \nu\LSP$ over $l^\pm\nu\LSP$, 
and why $\chi^0_2$ may decay to $\tau^+\tau^-\LSP$ over
$l^+l^-\LSP$.  Firstly,
$\stau$ is lighter than $\tilde l_R$.  If the $\tau$ Yukawa coupling
is large then renormalization group effects will drive $\stauR$ below
$\tilde l_R$. More importantly the $\tilde \tau$ mass matrix can have
a large mixing from $\lambda_\tau v\mu$, where $v=174\gev$.  This
mixing angle will drive the lightest eigenvalue down, thus potentially
making $m_\stau \ll m_{\tilde l_R}$.  This is especially true for large
$\tan\beta$ which enhances the $\tau$ Yukawa coupling,
\beq
\lambda_\tau \simeq \frac{m_\tau\tan\beta}{v} .
\eeq
The large $\tan\beta$ correlation with high $\tau$ multiplicity
has been recently emphasized in ref.~\cite{hightanbeta}.
Large mixing also results in models with large $A_\tau$ triscalar
soft mass Yukawa.  

If the slepton masses are lighter than $\chi^\pm_1$ and $\chi^0_2$
masses, then on-shell decays into $\stau$ will be more likely
than $\tilde l_R$.  Not only is the phase space larger for $\stau$, but
also the substantially mixing introduces a large $\tilde \tau_L$
component to $\stau$ thus coupling much more effectively to 
$\chi^\pm_1$ and $\chi^0_2$ which can be shown to be
mostly $SU(2)_L$ gauginos after radiative electroweak symmetry
breaking conditions are imposed in minimal 
supergravity~\cite{sugra} and
minimal gauge mediated models~\cite{mgm,babu,pierce}.  
(The result is more robust than
these minimal models.)
Furthermore, the large $\tau$ Yukawa coupling enables 
interactions with Higgsinos, which is negligible in the case of
$\tilde e$ or $\tilde \mu$.

Another reason why the $\tau$ final states might dominate can
be extracted from the ideas of ``more minimal'' 
supersymmetry~\cite{more minimal}.
According to this approach, all the first two generations squarks
and sleptons must be very heavy (greater than a few TeV)
to suppress large unwanted CP violation
and flavor changing neutral current effects in the electric dipole moment
of the neutron, $K -\bar K$ mixing, $\mu\to e\gamma$, etc.  The
$\tilde t_i$, $\tilde b_L$, Higgsinos and $SU(2)_L\times U(1)_Y$ must
be light in order to have natural electroweak symmetry breaking.
The remaining fields ($\tilde g$, $b_R$ and $\tilde \tau_i$) can be
either light or heavy without causing problems.  It is natural
to assume that they are light since $\tilde g$ is probably tied up
with the other gauginos in some unification relation, and it is
reasonable to put all scalars of the same generation at the
same scale.  Since $\tilde t_i$ and $\tilde b_L$ must be light, 
then by this argument so should the 
other third generation scalars $\tilde b_R$
and $\tilde \tau_i$.  In this case, decays of
$\chi^\pm_1$ and $\chi^0_2$ may only be allowed to decay into
$\tau$'s via $\stau$ since no other slepton is close in mass.
Although this approach may lead to new problems~\cite{heavy problems},
it is an attractive contributing solution to FCNC and CP violation suppression.

For minimal supergravity (mSUGRA) and minimal gauge mediated supersymmetry
breaking (mGMSB) models, we can quantify the prevalence of $3\tau$
events in the data with a small number of parameters.  For mSUGRA,
I have chosen $m_0=150\gev$, $m_{1/2}=225\gev$, $A_0=0$,
and ${\rm sign}(\mu)=+$ to be fixed for illustration 
purposes~\cite{explain sugra}.
Both $m_{\chi^\pm_1}$ and $m_{\chi^0_2}$ are close to $170\gev$ 
for all $\tan\beta$ and are mostly the superpartners of $W^\pm$
and $W^3$ gauge bosons respectively.
In Fig.~\ref{sugra_tau} the value of $\tan\beta$ is varied from
2 to 30 and the 
branching fraction of $3\tau$ events expected from $\chi^\pm_1\chi^0_2$
production and subsequent decays is plotted~\cite{isajet}.  
\jfig{sugra_tau}{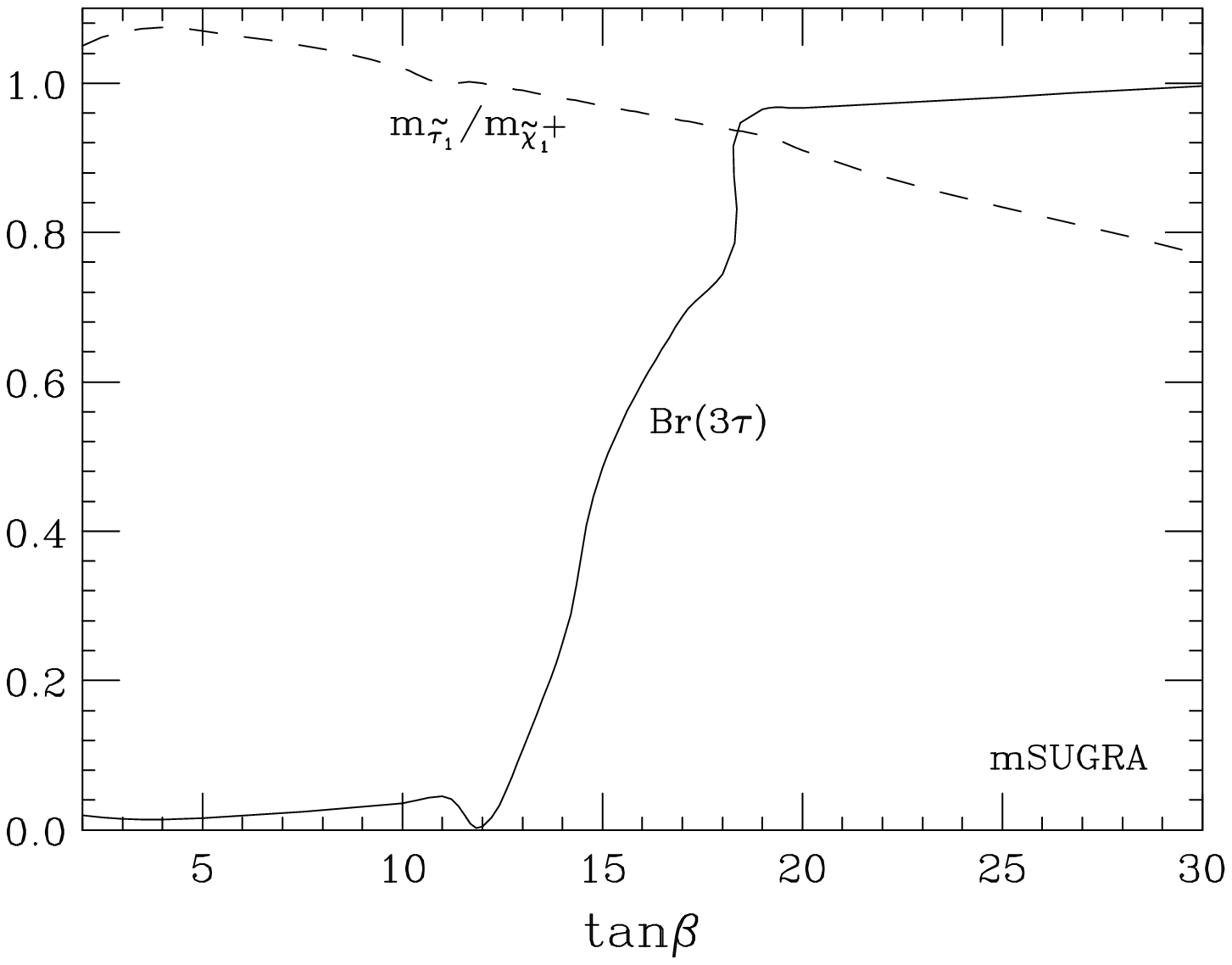}{Branching ratio of $\chi^\pm_1\chi^0_2$
into three $\tau$-leptons for the minimal supergravity model
described in the text.}
Furthermore, the
ratio of $m_{\stau}/m_{\chi^\pm_1}$ is shown.  As expected, the
$3\tau$ rate becomes dramatic for larger $\tan\beta$ and near almost
$100\%$ probability for $\tan\beta > 20$, when the $\stau$ mass dips
sufficiently below $\chi^+_1$.

In minimal gauge mediated supersymmetry (mGMSB) I have chosen
$\Lambda = M= 72\tev$, one $5+\bar 5$ messenger multiplet,
and ${\rm sign}(\mu)=+$~\cite{explain mgm}.  
Again, $m_{\chi^\pm_1}$ and $m_{\chi^0_2}$ are close to $170\gev$ 
for all $\tan\beta$ and are mostly the superpartners of $W^\pm$
and $W^3$ gauge bosons respectively.  We see in Fig.~\ref{mgm_tau}
that the ${\rm Br}(3\tau)$
branching rate turns on much more smoothly.  
\jfig{mgm_tau}{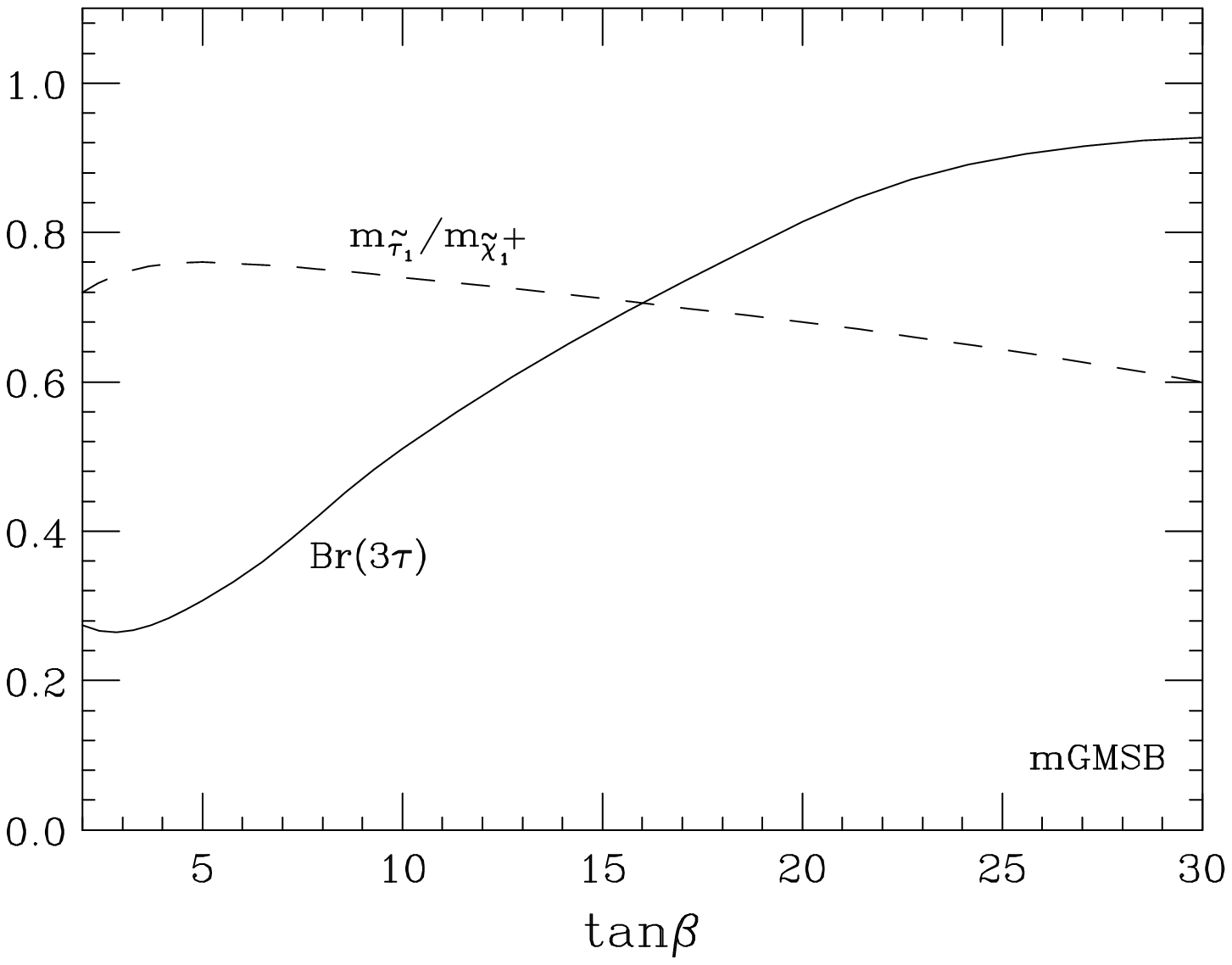}{Branching ratio of $\chi^\pm_1\chi^0_2$
into three $\tau$-leptons in the gauge mediated model
described in the text.}
Here, the spectrum
allows very light $\stau$ and $\tilde l_R$.  The dominance of
the $3\tau$ final state is mostly due to the increasing $\tilde \tau_L$
component of $\stau$ as $\tan\beta$ increases.  When $\tan\beta > 25$
the $3\tau$ branching fraction climbs above $90\%$.
Even if the decays of $\chi^\pm_1\to W^\pm\chi^0_1$ and
$\chi^0_2\to Z\chi^0_1$ are on-shell, they do not compete
with on-shell $\stau$ decay modes in large $\tan\beta$.

The $3\tau$ final state is most expected when $\tan\beta$ is
rather large.  Many grand unified theories based on
$b-\tau -t$ unification~\cite{b-tau-top unification} 
prefer large $\tan\beta \simeq m_t/m_b$.  In gauge mediated
models, unwanted large CP violation effects are suppressed
by requiring $B=0$ at the messenger scale~\cite{dinenelson,babu,mgm,rattazzi}.
This also implies large
$\tan\beta$, hence high $\tau$-multiplicity.  Therefore,
it is worthwhile to examine how effectively one can search
for $3\tau$ final states. 

\section{Searching in the $3\tau$ mode}
\medskip

Analysis for $\tau$ lepton final states is possible for both
CDF and D0 in the subsequent runs.  
The $\tau$ identification rate might be as high as 30\% in the hadron
decay modes
when $p_T(\tau)>15\gev$~\cite{private}.  
This is encouraging for analysis
of final state $\tau$ lepton signatures of new physics,
such as the $3\tau$ signature being considered here.
The main background for $3\tau$ events is $WZ\to 3\tau$ and 
$Zg$ where $Z\to \tau\tau$ and the $g$ fakes a $\tau$ with
the right jet-charge and low hadron multiplicity.  
The fake rate for this is expected to be approximately $1\%$
or less with $p_T>15\gev$.

One unifying approach to all trilepton signals, including $3l$
and $3\tau$, is to define an ``$\emt$ candidate.''  An $\emt$ candidate
is defined to be an isolated electron or muon or a fully
tagged $\tau$-jet in a hadronic decay mode of the $\tau$.  
Therefore, the identification of a primary $e$ or $\mu$ as
an $\emt$ candidate is near 100\%.  The identification of
a primary $\tau$ lepton is the sum of its branching ratios into
$e\nu\nu$ (18\%) and $\mu\nu\nu$ (18\%) plus the $\tau$ identification
rate (30\%) that utilizes the hadronic decay modes of the $\tau$.
The combined $\emt$ candidate efficiency is then 
approximately 66\% for a $\tau$ lepton.  
The estimate here is made assuming $p_T (\tau)>15\gev$,
and $\eta (\tau) < 1$.
For three $\tau$'s in an event satisfying these $p_T$ and $\eta$ requirements, 
the identification rate is as high as 
$(0.66)^3\simeq 30\%$.

The visible decay products of a $\tau$ can be significantly softer in
$p_T$ than the original $\tau$ itself.  For this reason, it is best to
insist that all three $p_T(\tau)$ in the signal
events are greater than $15\gev$ when analyzing search 
capabilities~\cite{pythia}.
In the end, a detailed detector simulation with well-defined $\tau$
identification requirements for each decay mode of the $\tau$ will be required
for complete understanding of the capabilities.  Here, I will approximate
this process by making sure that one $\tau$ satisfies
$p_T>20\gev$, and the other two $\tau$'s satisfy $p_T(\tau)>15$.
The trigger could occur by the substantial
missing $E_T$ in the events, and/or large enough $p_T$ of an isolated
lepton(s), and/or a dedicated $\tau$ trigger.
The trigger issues at both collaborations have not been settled but
the hope here is that 
requiring $p_T(\tau)=\{20,15,15\} \gev$ and $\eta(\tau)<\{ 1,1,1\}$
will yield a high overall trigger efficiency.

To reduce the $WZ$ background all events with two opposite sign same-flavor
leptons should satisfy $|m_{l^+l^-}-m_Z|>10\gev$.  
A small reduction in
$Zg$ can be obtained by cutting against back-to-back $\tau$'s in the
azimuthal plan; however, the high $p_T$ requirements on the ``$\tau$'' imposed
above make this background reduction rather insignificant.
Table~\ref{bkgtable} 
contains the estimated background to three $\emt$ candidate events
after cuts. 
\begin{table}
\centering
\begin{tabular}{cccccc}
\hline\hline
 & $\sigma$ [pb] & Br(``$3\tau$'') & $\epsilon_{\rm kin}$ &
    $\epsilon_{\rm id}$ &    $\sigma_{\rm cuts}$ [fb] \\
\hline
$W^\pm Z$ & 2.6 & 0.01 & 0.11 & 0.40 & 1.1 \\
$Zg$ & 8200 & $3.3\times 10^{-4}$ & 0.005 & 0.45 & 6.1 \\
\hline
total &  & & & & 7.2 fb \\
\hline\hline
\end{tabular}
\label{bkgtable}
\caption{Main backgrounds to $3\tau$ events at $2\tev$ center of mass energy.  
The Br(``$3\tau$'') indicates the
branching fraction into three $\emt$ candidates, but not allowing
$Z\to ee,\mu\mu$. The $\epsilon_{\rm kin}$ column is the efficiency of
selecting all three $\emt$ candidates with $p_T=\{ 20,15,15\}\gev$
and $\eta <1$.
The $\epsilon_{\rm id}$ column is the estimated probability of
experimentally identifying
all three leptons in the final state 
as $\emt$ candidates after the kinematic cuts have been applied. 
Note that the $WZ$ background mainly sums over $e\tau\tau$, $\mu\tau\tau$,
and $\tau\tau\tau$ final states, and the $\epsilon_{\rm id}$ averages
over these final states.  Also, the fake rate for $g\to \tau$ is incorporated
in Br(``$3\tau$'') not $\epsilon_{\rm id}$.}
\end{table}
A veto on additional jet activity above $E_T>15\gev$ has been applied
to help reduce additional background.  
For example, the $gg\to b\bar b$ background
rate is expected to be small with an effective veto in place.

To estimate the maximum supersymmetry mass reach for the $3\tau$ mode, the
following choices should be made: The $3\tau$ branching ratio is 100\% which is
valid for high $\tan\beta$.  The light charginos and neutralinos are mostly
gaugino so that $m_{\chi^\pm_1}=m_{\chi^0_2}=2m_{\chi^0_1}$, which is generally
true in supersymmetric models with radiative electroweak symmetry breaking, and
is certainly valid for mSUGRA and mGMSB.  And lastly, for maximum kinematic
efficiency we assume that $m_{\stau}$ is halfway between $m_{\chi^0_1}$
and $m_{\chi^0_2}$.  With these choices, the total signal rate after cuts
is 
\beq
\sigma_{\rm cuts}(3\tau~{\rm signal})\simeq 0.04\sigma(\chi^0_2\chi^\pm_1).
\eeq
Combing this equation with a background rate of $7.2\xfb$, it is required
that
\beq
\sigma (\chi^\pm_1\chi^0_2) \gsim \frac{330\xfb}
{\sqrt{\int dt{\cal L}~{\rm in}\xfb^{-1}}}
\eeq
to get a $5\sigma$ signal above background
over the applicable range for $m_{\chi^+_1}$.
Table~\ref{reach} indicates what the corresponding mass reach is for
these events for different integrated luminosity at $2\tev$ center of
mass energy.

The background and signal estimates require further consideration, 
especially after a detailed study has been completed on the 
possibility of triggering on $\tau$ leptons.  The results above
depend on having high efficiency triggers for central $\tau$ leptons
with $p_T\gsim 20\gev$. 
Because the leptonic decay products of the $\tau$ lepton are usually
quite soft, 
mass reaches of the chargino depend very sensitively on the trigger
capabilities of high $\tau$-multiplicity events~\cite{hightanbeta}.
\begin{table}
\centering
\begin{tabular}{cc}
\hline\hline
$\int dt {\cal L}$ & $m_{\chi^\pm_1}$ reach \\
\hline
$1\xfb^{-1}$ & $135\gev$ \\
2 & 145 \\
10 & 170\\
25 & 185 \\
\hline\hline
\end{tabular}
\label{reach}
\caption{Estimated maximum mass reach for $\chi^\pm_1$ in the
$3\tau$ channel of $\chi^\pm_1\chi^0_2$ production and decay.}
\end{table}

\section{Conclusion}
\medskip

The use of $\tau$ leptons in supersymmetry goes well beyond the
$3\tau$ signal discussed here.  High $\tau$ multiplicity
events are present
in some gauge mediated models when the $\stau$ is the 
NLSP~\cite{mgm,martin,nandi}.  In this
case two prompt decays of $\stau\to \tau +\tilde G$ may be present
in all superpartner events.  Without these $\tau$ leptons in the final state
the supersymmetry events
may be quite ordinary, and mimicked by large standard model backgrounds.

Tau leptons can also be used to gain significance in supersymmetry 
Higgs physics
observables.  
For example, many studies of Higgs boson
detection at the
Tevatron require $Wh\to l\nu b\bar b$, where $l=e$ or $\mu$.  Identifying
$\tau$ leptons in the decays 
of the $W$ and/or $h$~\cite{mrennakane,dreeshiggs} can
only help in the search for the Higgs boson.  
Since supersymmetry generally requires a light Higgs
boson with mass less than about $130\gev$, the Tevatron is well-poised
to discover it.  This requires high luminosity~\cite{tevhiggs} and taking
advantage of all possible production and decay modes of the Higgs. 

Nevertheless, the search reach of the pure $3\tau$ signal of supersymmetry
is a good benchmark for $\tau$ studies.  Not only is this final state
ubiquitous in highly motivated large $\tan\beta$ models of supergravity
and gauge mediated supersymmetry breaking frameworks, but it also
may be the only way to see supersymmetry in the more minimal supersymmetry
approaches where the $\tilde\tau$ is the only light slepton generation
in the spectrum.  

\bigskip
\noindent
{\it Acknowledgements:} I thank  D.~Chakraborty
and S.~Martin for helpful comments.



\begin{thebibliography}{20}

\bibitem{dinenelson}
M.~Dine, A.~Nelson, Y.~Shirman, \PRD{51}{95}{1362};
M.~Dine, A.~Nelson, Y.~Nir, Y.~Shirman, \PRD{53}{96}{2658}.

\bibitem{prompt}
For models allowing prompt decays see,
K.~Intriligator, S.~Thomas, \NPB{473}{96}{121}.

\bibitem{trileptons}
H.~Baer, K.~Hagiwara, X.~Tata, \PRD{35}{87}{1598};
P.~Nath, R.~Arnowitt, \MPL{2}{87}{331};
R.~Barbieri, F.~Caravaglios, M.~Frigeni, M.~Mangano, \NPB{367}{91}{28}.

\bibitem{tev_trilepton}
D0 Collaboration (B. Abbott et al.), \PRL{80}{98}{1591};
CDF collaboration (F. Abe et al.), hep-ex/9803015.

\bibitem{tevupgrade}
H.~Baer, C.-h.~Chen, C.~Kao, X.~Tata, \PRD{52}{95}{1565};
S.~Mrenna, G.L.~Kane, G.~Kribs, J.~Wells, \PRD{53}{96}{1168}.

\bibitem{hightanbeta}
H.~Baer, C.-h.~Chen, M.~Drees, F.~Paige, X.~Tata, \PRL{79}{97}{986};
hep-ph/9802441.

\bibitem{sugra}
M.~Drees, M.~Nojiri, \NPB{369}{92}{54};
V.~Barger, M.~Berger, P.~Ohmann, \PRD{49}{94}{4908};
G.L.~Kane, C.~Kolda, L.~Roszkowski, J.D.~Wells, \PRD{49}{94}{6173}.

\bibitem{mgm}
S.~Dimopoulos, S.~Thomas, J.~Wells, \NPB{488}{97}{39}.

\bibitem{babu}
K.~Babu, C.~Kolda, F.~Wilczek, \PRL{77}{96}{3070}.

\bibitem{pierce}
J.~Bagger, K.~Matchev, D.~Pierce, R.-J.~Zhang, \PRD{55}{97}{3188}.

\bibitem{more minimal}
A.~Cohen, D.~Kaplan, A.~Nelson, \PLB{388}{96}{588};
S.~Dimopoulos, G.~Giudice, \PLB{357}{95}{573}.

\bibitem{heavy problems}
N.~Arkani-Hamed, H.~Murayama, \PRD{56}{97}{6733};
K.~Agashe, M.~Graesser, hep-ph/9801446.



\bibitem{explain sugra}
See, for example, ref.~\cite{sugra} for an
explanation of the minimal supergravity parameters.

\bibitem{isajet}
The signal events were simulated using ISAJET: 
F.~Paige, S.~Protopopescu, in {\it Supercollider Physics}, ed. D.~Soper
(World Scientific, 1986); H.~Baer, F.~Paige, S.~Protopopescu,
X.~Tata, in {\it Proceedings of the Workshop on Physics at Current Accelerators
and Supercolliders}, ed. J.~Hewett, A.~White, D.~Zeppenfeld
(Argonne National Laboratory, 1993) hep-ph/9305342.

\bibitem{explain mgm}
See, for example, ref.~\cite{mgm} for an explanation of
the minimal gauge mediated parameters.

\bibitem{b-tau-top unification}
L.~Hall, R.~Rattazzi, and U.~Sarid, \PRD{50}{94}{7048};
G.~Anderson, S.~Dimopoulos, L.~Hall, S.~Raby, \PRD{47}{93}{3702}.

\bibitem{rattazzi}
R.~Rattazzi, U.~Sarid, \NPB{501}{97}{297}.


\bibitem{private}
D.~Chakraborty, private communication.

\bibitem{pythia}
Backgrounds are simulated using PYTHIA, T.~Sj\"ostrand, Comp.~Phys.~Comm.~
{\bf 82} (1994) 74.

\bibitem{martin}
S. Ambrosanio, G.L. Kane, G. Kribs, S.~Martin, S.~Mrenna,
\PRD{54}{96}{5395}; S.~Ambrosanio, G.~Kribs, S.~Martin,
hep-ph/9710217.

\bibitem{nandi}
B.~Dutta, S.~Nandi, hep-ph/9709511.

\bibitem{mrennakane}
S.~Mrenna, G.~Kane, hep-ph/9406337.

\bibitem{dreeshiggs}
For other $\tau$ observables in extended Higgs sector searches, see
M.~Drees, M.~Guchait, P.~Roy, \PRL{80}{98}{2047}.

\bibitem{tevhiggs}
A.~Stange, W.~Marciano, S.~Willenbrock, \PRD{50}{94}{4491};
S.~Kim, S.~Kuhlmann, W.-M.~Yao, ``Improvement of Signal 
Significance in $Wh\to l\nu b\bar b$ Search at TeV33,''
in ``Proceedings of the 1996 DPF/DPB Summer Study on New Directions
for High Energy Physics'' (1996);
S.~Mrenna, ANL-HEP-PR-97-35 (1997).














\end{thebibliography}
\end{document}